# Manganese nanoclusters and nanowires on GaAs surfaces

Mogus Mochena
Department of Physics, Florida A & M University, Tallahassee, Florida 32307

P. J. Lin-Chung
Naval Research laboratory, Washington, D. C. 20375

We have computed the local magnetic moments of manganese and neighboring arsenic for various cluster configurations on the (001) surface of GaAs bulk crystal using a cluster of 512 atoms. We obtained for manganese a substantial local magnetic moment of $3.66 \pm .01$ $\mu_B$ for all cases considered. The induced magnetic moment of arsenic is less than that of manganese by two orders of magnitude and falls off drastically beyond nearest neighbor distance. A small amount of charge is transferred from the manganese to arsenic atom. The possibility of a spin polarized wire channel on the arsenic layer below the surface is suggested.

# I. Introduction

There is a great deal of interest currently in introducing spin dependent functionality such as memory and storage capabilities into nonmagnetic semiconductors that have been widely used in technology such as silicon, germanium and galium arsenide [1, 2]. This trend is part of an emerging field that is commonly referred to as spintronics, the study of the possibility of utilizing the spin degree of freedom of an electron for device purposes. Semiconductor physics based on the manipulation of the electronic charge has been well established over the past four or five decades. Introducing such a new dimension to devices is a particularly interesting extension in light of current interest in quantum computing and quantum information theory [3].

Most of the current research towards realizing spin dependent semiconductor devices is focused in two areas: injecting spin polarized electrons from ferromagnets [4-7] or diluted magnetic semiconductors [8-10] into a semiconductor, or doping a semiconductor with a magnetic impurity [11-13]. Both efforts involve dealing with the physics of the bulk semiconductor. The behavior of the surfaces of these systems has received relatively little effort to date, except for the interface problem of spin injection.

Manganese clusters consisting of two to five atoms in free space have been shown in theoretical studies to retain their atomic magnetic moments, in sharp contrast to bulk manganese behavior [14]. Low dimensional systems such as surfaces and multilayers are well known to enhance magnetization [15]. The present work explores whether claims of large magnetic moments for clusters in [14] could be realized on a surface of a semiconductor. We have studied the effect of substituting manganese atoms at the sites of Ga on (001) surface of GaAs. A tight-binding model is used to take into account the interaction between the bulk semiconductor and the magnetic impurities. The choice of manganese and GaAs is based on their widespread use for study of diluted magnetic semiconductors.

Experimentally controlled deposition of atoms on semiconductor surface is easily realized with tunneling microscope [16]. We are not aware of any experiments involving such a deposition of few atoms of manganese on the surface of a semiconductor. The results of the present work, therefore, could be helpful to experimental investigations of induced magnetization on semiconductor surfaces.

The remainder of the paper is organized as follows. In section II, we briefly discuss the computational method that has been used to studying local effects, especially of large systems. In section III, we present our results for a dimer, a rectangular cluster, a wire, and a ladder consisting of two wires.

## II. Computational Method

We use a continuous fraction method that has been developed to study a local perturbation resulting from a defect, adatoms on surface, or local changes in a ferromagnetic or paramagnetic material [17,18]. The advantages of this method relative to the psuedopotential method, the small-cluster augmented method, the scattering theoretical method and the self-consistent Green's function technique have been discussed elsewhere [18]. The continued fraction method is based on the real space Green function which can be expanded as

$$G_{\alpha l \alpha l}(E) = \langle \alpha R_l | [E - H]^{-1} | \alpha R_l \rangle$$

$$= \cfrac{1}{E - a_0 - \cfrac{b_1^2}{E - a_1 - \cfrac{b_2^2}{E - a_2 - \cfrac{b_3^2}{E - .....}}}} \qquad (1)$$

where H is a tight binding Hamiltonian acting on localized orbital $\alpha$ at site $R_l$, $|\alpha R_l\rangle$. The coefficients $a_i$ and $b_i$ in the Green function are elements of a triadiagonal matrix representation of a tight binding Hamiltonian resulting from a unitary transformation on the local orbital basis $|\alpha R_l\rangle$ such that

$$\langle U_m | H | U_n \rangle = \begin{cases} a_m, & n = m \\ b_{m+1}, & n = m+1 \\ b_m^*, & n = m-1 \\ 0, & otherwise \end{cases} \qquad (2)$$

where

$$|U_0\rangle \equiv |\alpha R_l\rangle \qquad (3a)$$

$$b_n |U_n\rangle = (H - a_{n-1})|U_{n-1}\rangle - b_{n-1}|U_{n-2}\rangle \qquad (3b)$$

$$\langle U_n | U_n \rangle = 1 \qquad (3c)$$

A tridiagonal Hamiltonian representation has also been used to study elementary excitations at surfaces and interfaces [19].

As seen from the Green function expansion above, a complete knowledge of the coefficients is equivalent to determination of the Green function. The expansion in equation (1) is terminated when the convergence of results for a given size of cluster is achieved. Following [18] for the bulk GaAs case, we terminated the expansion at twenty five terms in this computation. The local density of states (LDOS), $N_{\alpha l}$, and the integrated DOS, $n_{\alpha l}$, are then given by

$$N_{\alpha l}(E) = -\pi^{-1} \operatorname{Im} G_{\alpha l \alpha l}(E + i\varepsilon) \qquad (4a)$$

$$n_{\alpha l} = \int_{-\infty}^{E_F} N_{\alpha l}(E) dE \qquad (4b)$$

From the integrated DOS, we obtain the magnetic moments and the amount of charge transfer between Mn and its arsenic neighbors.

In practice the coefficients are obtained by first determining the Slater-Koster (SK) parameters through an interpolative scheme from band structure calculations [18, 20]. In this work we used SK two-center parameters from earlier work [20-22] to construct a complete set of SK parameters in Table I. The parameters were scaled according to the prescription given by [23] to account for variations in lattice constants and according to [24] to account for coordination numbers. The energy contribution from the magnetic interaction is incorporated into the diagonal on-site energies as in the Hubbard model [25]. For Mn atoms, the 3d, 4s and 4p orbitals are used to study the mixing of the orbitals and the deviation of the magnetic moment from the free atom value. For the GaAs bulk crystal, which is represented as a large cluster of 512 atoms, only s and p orbitals are included. Currently there are no reliable SK parameters available for ferromagnetic Mn. Therefore, we used the paramagnetic values of Mn from [21] and modified their site energies according to the energy splitting for ferromagnetic iron as an approximation. The site energies are given in Table II.

III. Results and Discussion

We present first the results of substituting a single Mn at Ga site on (001) surface of the crystal to gain insight into local effects. We do not take into account surface reconstruction in this work; our goal is to get an understanding of the effect of the dangling bond on the magnetic interaction at an ideal surface. In Figure 1 we plot the LDOS for Mn substituted at the center of the surface for spin up and down states and similarly for a site at the center of the bulk for comparison. In Figure 2 the orbital DOS associated with Figure 1 are plotted. The d orbitals peak at their site energy of -1.73 eV for spin up and 0.52 eV for spin down states both for the surface and bulk states, and their structure remains more or less intact, in agreement with their localized nature. The s DOS is less broad and has a higher peak at the surface than at the center. The surface p DOS changes more from the center site both structurally and for the location of their peaks since they constitute higher energy levels than the other two orbitals. The overlap of s and p DOS suggests there is more mixing between the s and p orbitals with more interaction taking place at the center of the bulk than at the surface. The surface states are more localized because of the reduced degree of freedom and the dangling bonds. There is very little overlap between either the s or p orbitals with the d orbitals with a little bit more overlap for spin up s surface states.

Local magnetic moments are obtained from the difference of integrated DOS for spin up and down states at the Fermi energy which is determined by conservation of the total number of electrons locally. For a single Mn at a surface site, we fixed the Fermi energy by considering the total number of valence electrons on Mn and on the two nearest neighbor As sites. We obtained a magnetic moment of 3.65 $\mu_B$ for Mn and -0.013 for As.

This result seems reasonable in comparison to some of recent results of similar systems. The experimental magnetic moment of MnAs crystal is around 3.4 $\mu_B$, and Sanvito and Hill report a saturation magnetic moment of about 4 $\mu_B$ in their calculations [ 26, 27] by taking a pair of Mn and As sites. Our As moment is an order of magnitude smaller than that of Ref. 26. The amount of charge transfer from Mn to As is 0.096 e indicating some interaction between the neighbors. This flow of charge from Mn to As and a certain amount of mixing between the s and d orbitals at the Mn site could be responsible for the reduction of the Mn moment from the free atom value of $5\mu_B$. The 0.093 eV spin splitting of As bands gives an effective exchange coupling constant between the As and the Mn moment of J = -1.884 eV. This value compares reasonably well with similar values deduced from experiments for II-VI (-1.1eV ) [28] and for GaMnAs (-1.2 eV) [29]. Three higher values of 2.5 eV, 2.8 eV and 3.3 eV have also been reported [30, 31, 32].

Next we consider various cases to see the effect of substituting more Mn at Ga sites. Magnetic moments of thin films on semiconductor substrates are found to be negligible due to antiferromagnetic coupling between Mn [33]. Such coupling, at least in bulk case, according to theoretical arguments [34] depends on distance between the Mn sites; as distance decreases between the Mn, the coupling changes from ferromagnetic to antiferromagnetic. In diluted magnetic semiconductors, a ferromagnetic transition is observed at average Mn - Mn distance of 6 Å [32, 35]. According to recent studies of $Mn_xGe_{1-x}$, a ferromagnetic transition occurs at a distance of 10 Å [13]. Recently it has been reported that in small GaMnN clusters in free space Mn couples ferromagnetically with other Mn atoms, but antiferromagnetically with the nitrogen. The latter coupling in turn enhances the ferromagnetic coupling among the Mn atoms [36]. Therefore the remaining important issue is whether clusters of few Mn atoms or wire of Mn on a semiconductor surface, would also couple ferromagnetically.

We studied a dimer of Mn at the center of the surface, with Mn at next nearest neighbor distance from each other. The next nearest neighbor distance is the shortest distance between Mn in the substitutional case. Mn could also be placed at closer distances of interstitial or arsenic sites. In doping experiments, Mn prefers going into substitutional (Ga) and interstitial sites, but it rarely goes into As site [37]. We defer, however, the study of Mn at interstitial and arsenic sites that are closer than the next nearest neighbor distance to future work. The Fermi energy of the dimer is determined by the conservation of the total number of valence electrons among the dimer and three of the nearest As neighbors. One of the As ($As_1$) is bonded to both of the Mn of the dimer while the other two ($As_2$) are bonded singly to the Mn on either side. The moments are found to be 3.667 $\mu_b$ for Mn, -0.0211 for $As_1$ and -0.0151 for $As_2$. The Mn moment is the same as that of a single Mn impurity indicating the exchange interaction between Mn at next nearest neighbor distance is not strong enough to affect the moment. The spin polarization at the $As_1$ site is a factor of 1.6 greater than that for a single substitution Mn case, whereas that for $As_2$ deviates only slightly from the single Mn case. The spin polarization drops off drastically to -0.00028 $\mu_b$ at the As site that is a neighbor to $As_2$ on the same layer but on the opposite side to $As_1$.

Next we considered a rectangular cluster consisting of two dimers. The magnetic moment of Mn remained the same as that of a single dimer, confirming the short ranged nature of the exchange interaction. The spin polarization of As is same as that of a dimer and is limited to nearest neighbor distance from Mn as before.

We also investigated a wire of Mn atoms on the (001) GaAs surface layer. Then we considered two parallel wires forming a ladder on the surface. The magnetic moments of the Mn were the same in both cases as that of a single substituting impurity confirming once more the short-ranged nature of the exchange interaction. The spin polarizations of As in the two-dimer rectangle case and ladder case are similar to that of the single dimer case. Thus the ladder configuration may provide a spin polarized wire channel for the transport of holes on the layer beneath the surface.

Our results indicate that a substitutional monolayer of Mn on the surface of GaAs may also retain the ferromagnetic behavior. This indication is consistent with the recent experimental observations of high ferromagnetic transition temperature in Mn δ-doped GaAs heterostructure. [38]

The sizable local moments retained at Mn sites can be explained in terms of integrated DOS. In Figure 3, a typical relationship between integrated local DOS and energy is plotted. At high energies both spin up and down states are equally likely at a site so that the net magentic moment is zero. At low energies, the sites will be occupied by spin up states only, which is a case of total polarization. In the intermediate range where the spin down curve is rising in Figure 3, and where the Fermi energy lies in our calculation, both spin up and down states are probable. Since the spin up curve rises faster, and also earlier, it leads to a net moment.

In conclusion, our results clearly indicate that locally magnetized nanostructures are possible when magnetic impurities such as manganese are substituted into a tetrahederal structure on a semiconductors surface. Ferromagnetic coupling is possible at second nearest neighbor distance between Mn in GaAs and perhaps enhanced by the much smaller antiferromagnetic coupling to arsenic. While the moment of Mn is due to the localized d orbitals, the less localized but spin polarized orbitals of arsenic below a surface layer of Mn atoms could be used to create spin currents.

Acknowledgements

The authors are grateful to Dr. A. K. Rajagopal for suggesting the problem of nanoclusters on a surface of a semiconductor and many helpful discussions. MM is thankful to ASEE for support during the course of this work at NRL. PJLC is supported in part by the Office of Naval Research.

Table I

Slater - Koster parameters in Rydberg units. The superscripts stand for neighboring distances.

|     | Ga-As   | As-As[1] | As-As[2] | Ga-Ga   | Mn-Mn[1] | Mn-Mn[2] | As-Mn  | Ga-Mn   |
|-----|---------|----------|----------|---------|----------|----------|--------|---------|
| $ss\sigma$ | -0.1311 | -0.107   | -0.0011  | -0.0036 | -0.068   | -0.032   | -0.088 | -0.0178 |
| $sp\sigma$ | 0.1478  | 0.064    | 0.0016   | 0.0043  | 0.149    | 0.054    | 0.107  | 0.0291  |
| $pp\sigma$ | 0.3665  | 0.191    | 0.0492   | 0.0629  | 0.179    | 0.052    | 0.185  | 0.0578  |
| $pp\pi$ | -0.1833 | -.325    | -0.0248  | -0.0315 | 0.049    | 0.0012   | 0.0    | -0.0151 |
| $sd\sigma$ |         |          |          |         | -0.021   | -0.014   | -0.010 | -0.007  |
| $pd\sigma$ |         |          |          |         | -0.05    | -0.025   | -0.025 | 0.012   |
| $pd\pi$ |         |          |          |         | 0.023    | 0.0047   | 0.0115 | 0.0024  |
| $dd\sigma$ |         |          |          |         | -0.051   | -0.006   |        |         |
| $dd\pi$ |         |          |          |         | 0.042    | 0.0017   |        |         |
| $dd\delta$ |         |          |          |         | -0.0072  | -0.00002 |        |         |

Table II

Site energies of Mn in Rydberg at the surface, and at the center of GaAs crystal. The site energies for Ga and As are given in Ref. [20].

|    | Surface Up | Surface Down | Bulk Up | Bulk Down |
|----|------------|--------------|---------|-----------|
| 3d | -0.1269    | 0.0381       | -0.1329 | 0.0281    |
| 4s | 0.0896     | 0.1976       | 0.1616  | 0.1256    |
| 4p | 0.6426     | 0.7086       | 0.6161  | 0.7411    |

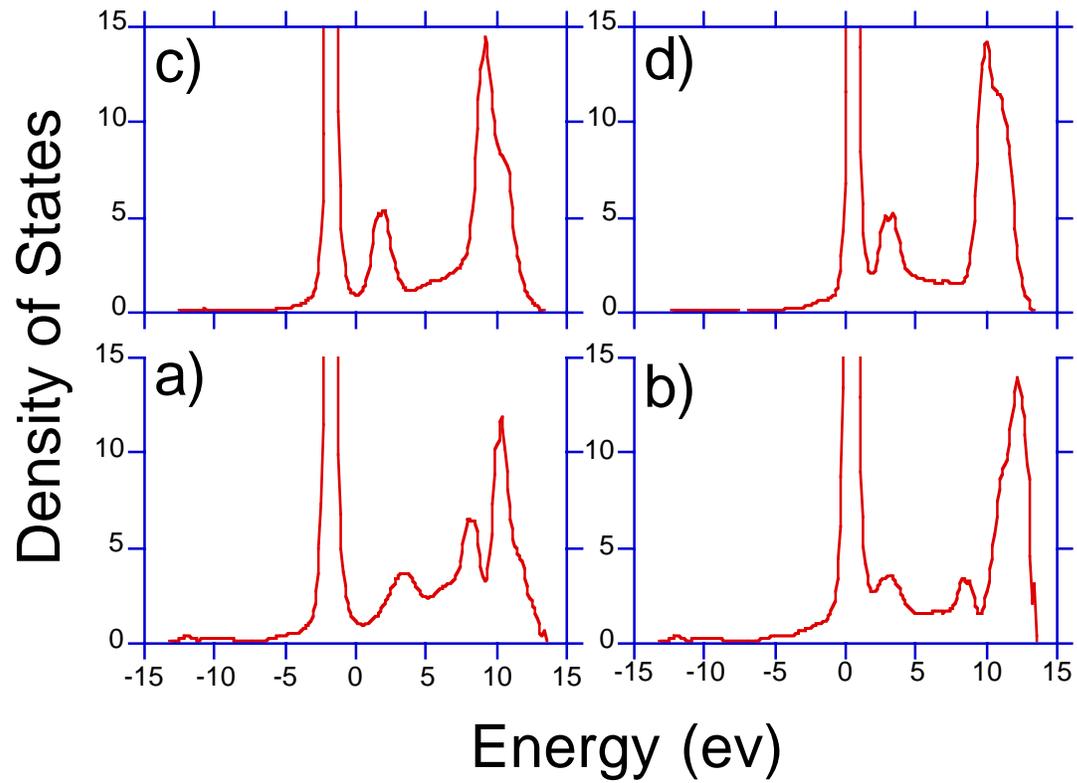

Fig. 1. Density of states for Mn at the center and surface of a 512 atoms cluster, a) and b) are for spin up and down states at the center, and c) and d) are for a surface site.

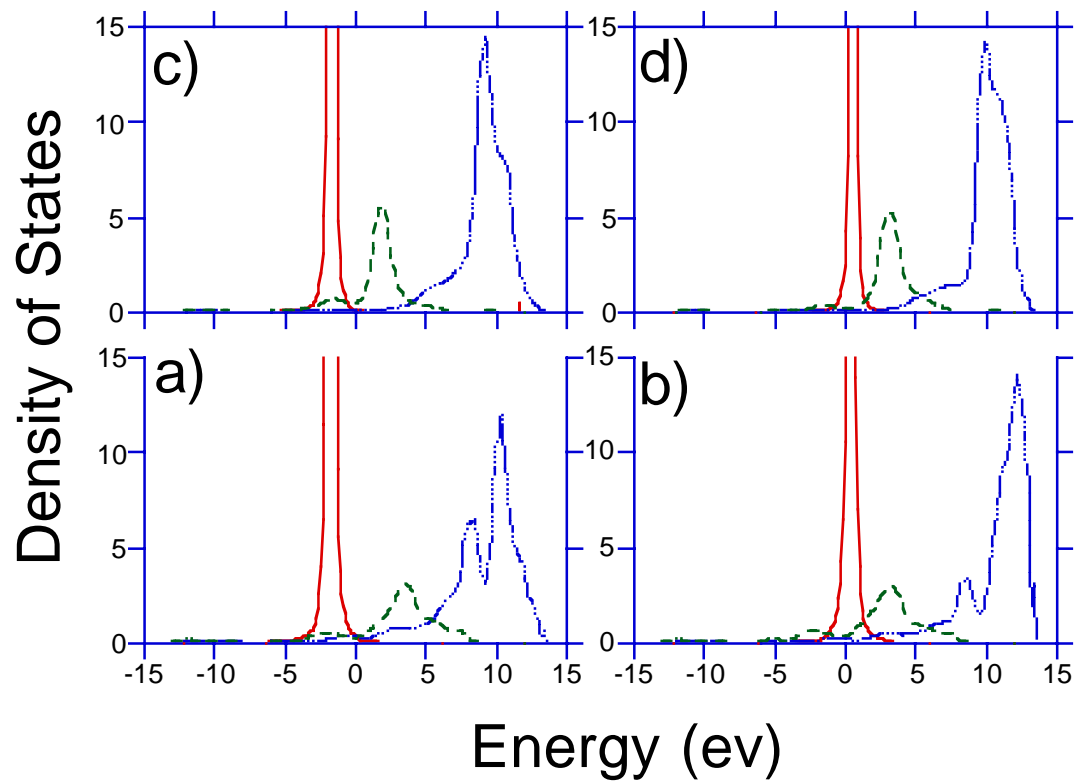

Fig. 2. Density of states for orbitals of Mn at the center and surface of a 512 atoms cluster, a) and b) are for spin up and down states at the center, and c) and d) are for a surface site. The solid lines are d, the dotted lines are p, and the dashed lines are s orbitals.

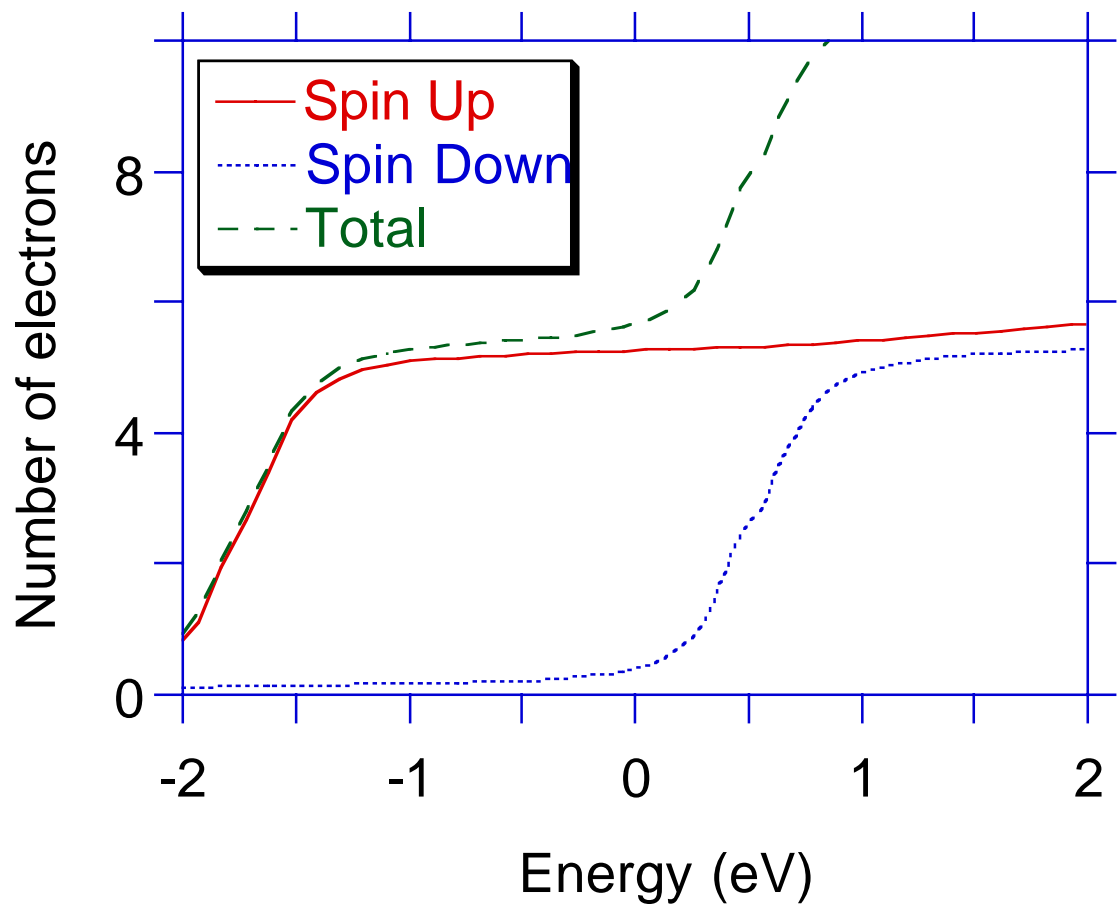

Fig. 3. Integrated density of states for Mn of a dimer.